\documentclass[final,5p,times,twocolumn,nopreprintline]{elsarticle}

\usepackage{amssymb}
\usepackage{amsmath}
\usepackage{booktabs}
\RequirePackage{flushend}

\usepackage{color}

\newcommand{\mD}{M_D}
\newcommand{\mpi}{M_\pi}
\newcommand{\mK}{M_K}
\newcommand{\F}{\mathcal{F}}
\newcommand{\N}{\mathcal{N}}
\newcommand{\M}{\mathcal{M}}
\newcommand{\nnnl}{\nonumber\\}
\newcommand{\ol}{\bar}
\newcommand{\diff}{\text{d}}
\newcommand{\disc}{{\rm disc}\,}
\newcommand{\FF}{\mathrm{FF}}

\newcommand{\beq}{\begin{equation}}
\newcommand{\eeq}{\end{equation}}

\begin{document}

\begin{frontmatter}

\title{Consistent Dalitz plot analysis of Cabibbo-favored $D^+\to \bar{K} \pi \pi^+$ decays}

\author[HISKP]{Franz Niecknig}
\author[HISKP]{Bastian Kubis}

\address{Helmholtz-Institut f\"ur Strahlen- und Kernphysik (Theorie) and
   Bethe Center for Theoretical Physics, Universit\"at Bonn, 53115 Bonn, Germany}

\begin{abstract} 
We resume the study of the Cabibbo-favored charmed-meson decays $D^+\to \bar{K} \pi \pi^+$ in a dispersive framework that satisfies unitarity, analyticity, and crossing symmetry by construction.
The formalism explicitly describes the strong final-state interactions between all three decay products and relies on pion--pion and pion--kaon phase shift input.
For the first time, we show that the $D^+\to K_S \pi^0 \pi^+$ Dalitz plot obtained by the BESIII collaboration as well as the $D^+\to K^- \pi^+ \pi^+$ Dalitz plot data by CLEO and FOCUS can
be described consistently, 
exploiting the isospin relation between the two coupled decay channels that provides better constraints on the subtraction constants.

\end{abstract}

\begin{keyword}
Hadronic decays of charmed mesons\sep Meson--meson interactions\sep Dispersion relations\sep Partial-wave analysis
\PACS 13.25.Ft\sep 13.75.Lb\sep 11.55.Fv\sep 11.80.Et
\end{keyword}

\end{frontmatter}


\section{Introduction}\label{sec:intro}
Three-body decays of heavy mesons provide a powerful mean for Standard Model tests and beyond. Due to their richer kinematic structure as compared to two-body decays, they have
a wide range of applications e.g.\ in hadron spectroscopy, studies of hadronic final-state interactions, or CP-violation studies.
For these investigations, a thorough understanding of the strong final-state interactions is mandatory,
necessitating the methods of amplitude analysis~\cite{Battaglieri:2014gca}.

One particular issue in this area are sensible parametrizations of scalar partial waves.  It is well known 
that the lowest-lying scalars cannot be described in terms of Breit--Wigner functions.  
In particular the lightest pion--pion and pion--kaon scalar resonances, the $f_0(500)$ (or $\sigma$)
and the $K_0^*(800)$ (or $\kappa$), are associated with poles in the complex energy plane too far away from 
the real axis to allow for any such simplistic 
description~\cite{Caprini:2005zr,GarciaMartin:2011jx,DescotesGenon:2006uk,Pelaez:2016klv}.
This has been recognized also in the context of heavy-meson decays, where more appropriate descriptions
in terms of scalar form factors have been applied~\cite{GardnerMeissner,Daub:2015xja,Albaladejo:2016mad},
using model-independent methods such as chiral perturbation theory and dispersive techniques.

A different aspect of three-body decays that requires refinement compared to a Breit--Wigner type isobar description
lies precisely in the presence of a third hadron in the final state: crossed-channel rescattering effects will necessarily modify the spectral forms of different resonances, and the extent to which this is the case (and can potentially be regarded as a ``small correction'') is too seldomly investigated explicitly.
One well-established theoretical tool to study such modifications are the
Khuri--Treiman equations~\cite{KhuriTreiman,Bronzan,Aitchison:1965zz,Aitchison:1966lpz,Pasquier:1968zz,Pasquier:1969dt}
(see also the recent lecture notes in Ref.~\cite{Aitchison:2015jxa}), 
originally invented to study $K\to3\pi$ decays in a manner consistent with the constraints from 
analyticity (i.e., causality) and unitarity (i.e., probability conservation).
They were resurrected and applied extensively to study 
$\eta\to 3\pi$~\cite{AnisovichLeutwyler,Kambor:1995yc,Guo:2015zqa,Guo:2016wsi,Colangelo:2016jmc,Albaladejo:2017hhj},
and have subsequently also been applied to other low-energy three-body decays such as 
$\omega/\phi\to 3\pi$~\cite{Niecknig:2012sj,Danilkin:2014cra}
or $\eta'\to\eta\pi\pi$~\cite{Schneider:2012nng,etaprime}.

Recently, we have applied Khuri--Treiman equations to the Cabibbo-favored charmed-meson
decays $D^+\to K^- \pi^+ \pi^+/\bar{K}^0 \pi^0 \pi^+$~\cite{Niecknig:2015ija}. 
In comparison to the light-meson decays mentioned above, the corresponding Dalitz plot is 
significantly larger, with a much richer spectrum of partial waves/resonances contributing.
In particular the decay $D^+\to K^- \pi^+ \pi^+$ has been theoretically studied frequently 
before~\cite{Oller:2004xm,Boito:2009qd,Magalhaes:2011sh,Magalhaes:2015fva,Guimaraes:2014cca}
(see also Refs.~\cite{ElBennich:2006yi,ElBennich:2009da} for analyses of similar $B$-meson decays),
using various approximations in the description of final-state interactions.
This corresponds to the
rather good data situation for that channel~\cite{Aitala:2005yh,FOCUS,CLEO,Link:2009ng}.
However, a consistent, combined investigation of both final states is highly desirable, 
as they are coupled to each other by simple charge-exchange rescattering, but only the partially neutral
final state allows for the observation of resonances in the pion--pion system (essentially the $\rho(770)$), 
while the $\pi^+\pi^+$ system is necessarily a nonresonant isospin $I=2$ state.
In simple isobar models as conventionally used in experimental analyses, neglecting the interaction of all three final-state particles, the relation between both channels is therefore obviously lost.
In addition to Ref.~\cite{Niecknig:2015ija}, such a combined theoretical analysis has only been performed
in Ref.~\cite{Nakamura:2015qga}, in the latter case using Faddeev equations to generate three-body rescattering
effects.  With the advent of experimental data on $D^+\to \bar{K}^0 \pi^0 \pi^+$, courtesy of the 
BESIII collaboration~\cite{Ablikim:2014cea}, we are now in the position for the first time to test our theoretical 
approach for consistency, using real data for both channels.  

This is the purpose of the current letter.
We briefly summarize the dispersion-theoretical formalism developed in Ref.~\cite{Niecknig:2015ija} 
in Sect.~\ref{sec:kinamp}, 
before performing fits to the new BESIII data as well as combined fits for both final states in Sect.~\ref{sec:compdata}.
We find the need to somewhat improve on the amplitude representation in particular with regards to the 
$D$-wave, which is discussed in Sect.~\ref{sec:DwaveOm}.  We conclude our study in Sect.~\ref{sec:conclusion}.

\section{Kinematics, decay amplitude, dispersive representation}\label{sec:kinamp}
We define the Mandelstam variables of the three-particle decays
\begin{equation}
D^+(p_D)\to \bar K(p_K)\,\pi(p_1)\,\pi^+(p_2)
\end{equation}
by $s=(p_{D}-p_1)^2$, $t=(p_{D}-p_2)^2$, and $u=(p_{D}-p_{K})^2$.
The corresponding crossed-channel scattering angles are given by
\begin{align}
&z_s \equiv\cos\theta_s =\frac{s(t-u)-\Delta}{\kappa(s)}\,,
\quad
z_t\equiv\cos\theta_t=\frac{t(s-u)-\Delta}{\kappa(t)}\,, \nnnl
& z_u\equiv\cos\theta_u=\frac{t-s}{\kappa_u(u)}\,,
\end{align}
with $\Delta=\big(\mD^2-\mpi^2\big)\big(\mK^2-\mpi^2\big)$ and
\begin{align}
&\kappa(x) =\lambda^{1/2}(x,\mK^2,\mpi^2)\lambda^{1/2}(x,\mD^2,\mpi^2)\,, \nnnl
&\kappa_u(u)=\lambda^{1/2}(u,\mD^2,\mK^2)\sqrt{1-\frac{4\mpi^2}{u}}\,,
\end{align} 
where the K\"all\'en function is defined by $\lambda(x,y,z)\equiv x^2+ y^2 +z^2-2(xy+xz+yz)$.
The decay amplitudes are decomposed into amplitudes depending on one Mandelstam variable only along the lines of the so-called reconstruction theorem~\cite{Kambor:1995yc,Stern:1993rg, Knecht:1995tr,Ananthanarayan:2000cp,Zdrahal:2008bd,Stoffer:2014rka}.
The explicit decompositions for the decay channels considered here have been performed in Ref.~\cite{Niecknig:2015ija} and read
\begin{align}
&\M_{\ol{0}0+}(s,t,u)=\frac{1}{2\sqrt{2}}\Big(\sqrt{3}(t-s)\,\F^1_1(u)-\F^2_0(u)\Big)\nnnl
&+\sqrt{\frac{3}{5}} \F^{3/2}_0(s) 
-\frac{2}{\sqrt{15}}\F^{3/2}_0(t)+\frac{1}{\sqrt{6}}\F^{1/2}_0(t)
\nnnl
&-\frac{1}{\sqrt{6}}\big[t(s-u)-\Delta\big]\,\F^{1/2}_1(t) \nnnl
&-\frac{1}{2\sqrt{6}}\Big[3\big(t(s-u)-\Delta\big)^2-\kappa^2(t)\Big]\,\F^{1/2}_2(t)
\,\label{eq:Full00p}
\end{align}
for the $D^+\to \bar{K}^0\pi^0\pi^+$ decay  and
\begin{align}
&\M_{-++}(s,t,u)=\F^2_0(u)
+\Bigg\{ \frac{1}{\sqrt{3}}\F^{1/2}_0(s) -\sqrt{\frac{2}{15}}\F^{3/2}_0(s) \nnnl
&
+\frac{1}{\sqrt{3}}\big[s(t-u)-\Delta\big]\,\F^{1/2}_1(s)\nnnl
&+\frac{1}{2\sqrt{3}}\Big[3\big(s(t-u)-\Delta\big)^2-\kappa^2(s)\Big]\,\F^{1/2}_2(s) + (s \leftrightarrow t) \Bigg\} \, \label{eq:Fullmmp}
\end{align}
for $D^+\to K^-\pi^+\pi^+$.
The strong final-state interactions of both decay channels are isospin-related and can therefore be described by the same single-variable amplitudes $\F^I_L$ of isospin $I$ and angular momentum $L$.
The above decomposition is consistently truncated beyond $D$-waves, and we have neglected exotic, nonresonant partial waves beyond the $S$-waves, i.e.\ the tiny $\pi K$ $P$- and $D$-wave amplitudes of isospin $I=3/2$ as well as the $\pi\pi$ $I=2$ $D$-wave.\footnote{Note that $F$-wave resonances in both the $\pi K$ and the $\pi\pi$ system are too heavy to contribute inside the Dalitz plot.}
The $\F^I_L$ satisfy the following elastic unitarity relations:
\beq\label{eq:unrel}
 \disc\F_L^I(x) = 2i\,\Big(\F_L^I(x)+\hat\F_L^I(x)\Big)\,\theta\,(x-x_\mathrm{th})\,\sin\delta_L^I(x)e^{-i\delta_L^I(x)}\,,
\eeq
where $x_\mathrm{th}$ denotes the elastic threshold of the channel considered, and $\delta^I_L(x)$ the corresponding $\pi\pi$/$\pi K$ scattering phase 
shift input of isospin $I$ and angular momentum $L$ taken from Refs.~\cite{Caprini:2011ky,CapriniWIP,GarciaMartin:2011cn,piK} (see also Ref.~\cite{Pelaez:2016tgi} for a new analysis of $\pi K$ scattering).
We attribute uncertainty bands to all phase shifts that rise linearly from zero at threshold (for the $S$-waves) or the position of the first resonance (for the $P$- and $D$-waves) to $\pm20^\circ$ at $2\,\text{GeV}$; see Fig.~\ref{fig:PWBES} (right column) below.
The inhomogeneities  $\hat{\F}_L^I(x)$ are given by the subsequent partial-wave projections
\begin{align}
\hat{\F}_L^I(x)&=\frac{2L+1}{2a_{I,L}^{ijk}\,\kappa^L_x(x)}\int_{-1}^1 \diff z_x\M_{ijk}^{I_x}(x,z_x)P_L(z_x)-\F_L^I(x)\,,\label{eq:inhomodet}	
\end{align}
and give rise to crossed-channel rescattering contributions. They are calculated explicitly for all channels in Ref.~\cite{Niecknig:2015ija}. The constants $a_{I,L}^{ijk}$ denote the Clebsch--Gordan coefficients corresponding to the single-variable amplitudes 
for the final states $ijk\in\{\bar00+,-++\}$.
The solutions to these unitarity relations, Eq.~\eqref{eq:unrel}, are given in the form of inhomogeneous Omn\`es solutions:
{\allowdisplaybreaks
\begin{align}
\F_0^2(u)&=\Omega^2_0(u)\frac{u^2}{\pi}\int_{u_\mathrm{th}}^\infty \frac{\diff u'}{u'^2}
 \frac{\hat{\F}_0^2(u')\sin\delta^2_0(u')}{\left|\Omega^2_0(u')\right|(u'-u)} \,,\nnnl
\F_1^1(u)&=\Omega^1_1(u)\Bigg\{c_0+c_1u+\frac{u^2}{\pi}\int_{u_\mathrm{th}}^\infty\frac{\diff u'}{u'^2}
 \frac{\hat{\F}_1^1(u')\sin\delta^1_1(u')}{\big|\Omega^1_1(u')\big|(u'-u)}\Bigg\}\,,\nnnl
\F_0^{1/2}(s)&=\Omega^{1/2}_0(s)\Bigg\{c_2+c_3s+c_4s^2+c_5s^3\nnnl
 &\qquad\qquad +\frac{s^4}{\pi}\int_{s_\mathrm{th}}^\infty\frac{\diff s'}{s'^4}
 \frac{\hat{\F}_0^{1/2}(s')\sin\delta^{1/2}_0(s')}{\big|\Omega^{1/2}_0(s')\big|(s'-s)}\Bigg\}\,,\nnnl  
\F_0^{3/2}(s)&=\Omega^{3/2}_0(s)\Bigg\{\frac{s^2}{\pi}\int_{s_\mathrm{th}}^\infty\frac{\diff s'}{s'^2}
 \frac{\hat{\F}_0^{3/2}(s')\sin\delta^{3/2}_0(s')}{\big|\Omega^{3/2}_0(s')\big|(s'-s)}\Bigg\}\,,\nnnl
 \F_1^{1/2}(s)&=\Omega^{1/2}_1(s)\Bigg\{c_6+\frac{s}{\pi}\int_{s_\mathrm{th}}^\infty\frac{\diff s'}{s'}
 \frac{\hat{\F}_1^{1/2}(s')\sin\delta^{1/2}_1(s')}{\big|\Omega^{1/2}_1(s')\big|(s'-s)}\Bigg\}\,,\nnnl  
 \F_2^{1/2}(s)&=\Omega^{1/2}_2(s)\frac{1}{\pi}\int_{s_\mathrm{th}}^\infty\diff s'
 \frac{\hat{\F}_2^{1/2}(s')\sin\delta^{1/2}_2(s')}{\big|\Omega^{1/2}_2(s')\big|(s'-s)}\,,\label{eq:fulleq}
\end{align}}%
with  $\Omega_L^I(x)$ the corresponding Omn\`es functions
\beq
\Omega_L^I(x)=\exp\biggl\{\frac{x}{\pi}\int_{x_\mathrm{ th}}^\infty \diff x'\frac{\delta_L^I(x')}{x'(x'-x)}\biggr\} \,.
\eeq
While built on the requirement of fulfilling two-body unitarity, Eq.~\eqref{eq:unrel}, the amplitude representation remarkably also fulfills the constraints of three-body unitarity~\cite{Aitchison:1966lpz,Aitchison:2015jxa} (compare also Ref.~\cite{Mai:2017vot}).
The $c_i$ denote the seven (complex) subtraction constants that are mandatory to obtain convergent dispersion integrals.
These subtraction constants cannot be determined by dispersion theory alone and have to be fixed by more fundamental dynamical theories or, as performed here, by fits to experimental data.
In particular, it might be possible to constrain their imaginary parts by three-body-unitarity considerations; this will be difficult in practice, however, as the three-body invariant mass is fixed to that of the decaying $D$-meson.
The constraint on the Mandelstam variables $s+t+u=\text{const.}$ has allowed us to eliminate some of the subtraction
constants, which are equivalent to the leading Taylor coefficients in an expansion around $s/t/u=0$, for some of the single-variable functions
in order to obtain a unique decomposition of the full decay amplitude.
We have chosen the nonresonant $I=3/2$ and $I=2$ amplitudes for that purpose; see Ref.~\cite{Niecknig:2015ija} for 
all details.
Strictly speaking, the single-variable amplitudes in the above decomposition need an even higher degree of subtractions due to the high-energy behavior of the $D$-wave $\F_2^{1/2}$.
The high-energy behavior of the decay amplitudes, and thus the single-variable amplitudes times their polynomial prefactors, is chosen to be consistent with the Froissart bound, 
which the $\F_2^{1/2}$ amplitude cannot satisfy.\footnote{For the assumed asymptotic behavior of the input phase shifts that determines the one of the Omn\`es functions and hence the single-variable amplitudes, we refer to Ref.~\cite{Niecknig:2015ija}.}
The thorough inclusion of the $\pi K$ isospin $1/2$ $D$-wave would thus necessitate more unknown subtraction constants that lower the predictability of the system.
We therefore choose to include the $\F_2^{1/2}$ amplitude heuristically in the sense that we exclude all crossed-channel projections of the $\F_2^{1/2}$ amplitude in the representation above: the other (lower) partial waves are allowed to contribute to the inhomogeneity $\hat\F_2^{1/2}$, but not vice versa.
For an in-depth derivation of the full decay amplitude as well as the single-variable amplitudes we refer the reader to Ref.~\cite{Niecknig:2015ija}.

The set of dispersion relations~\eqref{eq:fulleq} is linear in the single-variable amplitudes $\F_J^I(s)$ as well as in the subtraction constants. 
To solve the system it is therefore beneficial to exploit this linearity and construct a basis of the solution space.
The basis functions that span this solution space can be obtained by choosing a maximal set of linearly independent subtraction-constant configurations and solve the integral equations for each configuration.
We choose the subtraction-constant configuration $c_j=\delta_{ij},~j\in\{0,1,\ldots,n-1\}$ for the $i$th basis function $\M_i(s,t,u)$.  
Thus the general solution can be written as a linear combination
\beq
\M_{\bar0 0 +}(s,t,u)=\sum_{i=0}^{n-1} c_i \M_i(s,t,u)
\eeq
(and $n=7$ in the system~\eqref{eq:fulleq}).
The explicit numerical solution strategy to determine the basis functions via matrix inversion is discussed in detail in Ref.~\cite{Niecknig:2015ija}.  

\section{Experimental comparison}\label{sec:compdata}
In this section we perform fits of the dispersively determined decay amplitudes, displayed above, to the experimental $D^+\to K^0_S\pi^0\pi^+ / K^-\pi^+\pi^+$ data of the BESIII~\cite{Ablikim:2014cea}, CLEO~\cite{CLEO}, and FOCUS~\cite{FOCUS} collaborations.
The $D^+ \to K^0_S\pi^0\pi^+$ Dalitz plot is totally dominated by the $D^+\to \bar{K}^0\pi^0\pi^+$ decay, since the $D^+ \to K^0\pi^0\pi^+$ decay channel is doubly Cabibbo-suppressed.

Previously, the same amplitudes have been compared to the $D^+\to K^-\pi^+\pi^+$ Dalitz plot data in detail~\cite{Niecknig:2015ija}.
Here we will focus on the $D^+ \to K^0_S\pi^0\pi^+$ data from the BESIII collaboration and
subsequently perform combined fits to the $D^+\to K^0_S\pi^0\pi^+ / K^-\pi^+\pi^+$ data sets to use the isospin relation between these channels to full capacity, as well as to check the extracted subtraction constants for consistency. 
Furthermore we will discuss the inclusion of the $\pi K$ $D$-wave in more detail.

\subsection{Comparison to the BESIII data}\label{sec:BESdata}
\begin{figure}[t]
\centering
\includegraphics*[width=\linewidth]{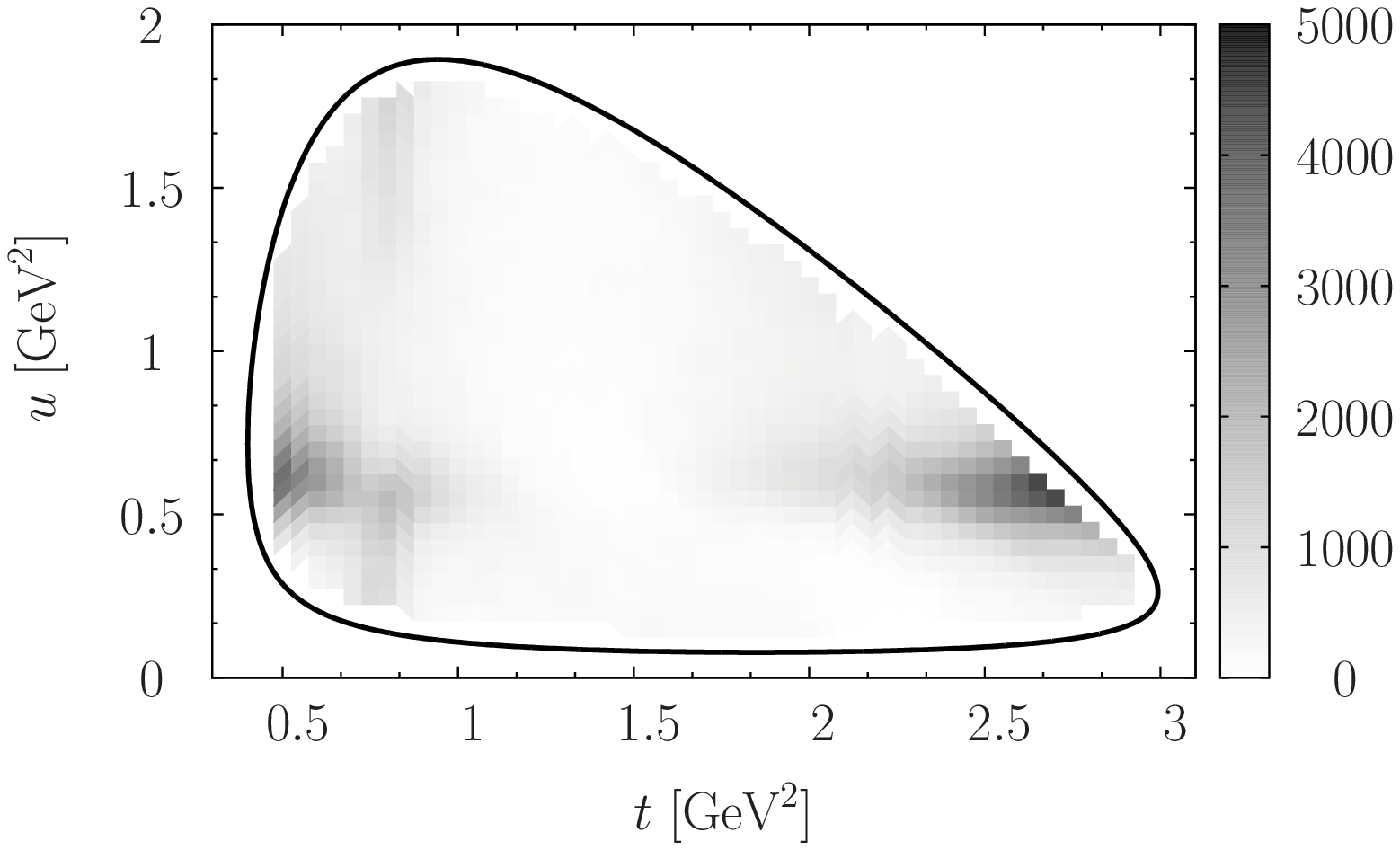}\\
\includegraphics*[width=\linewidth]{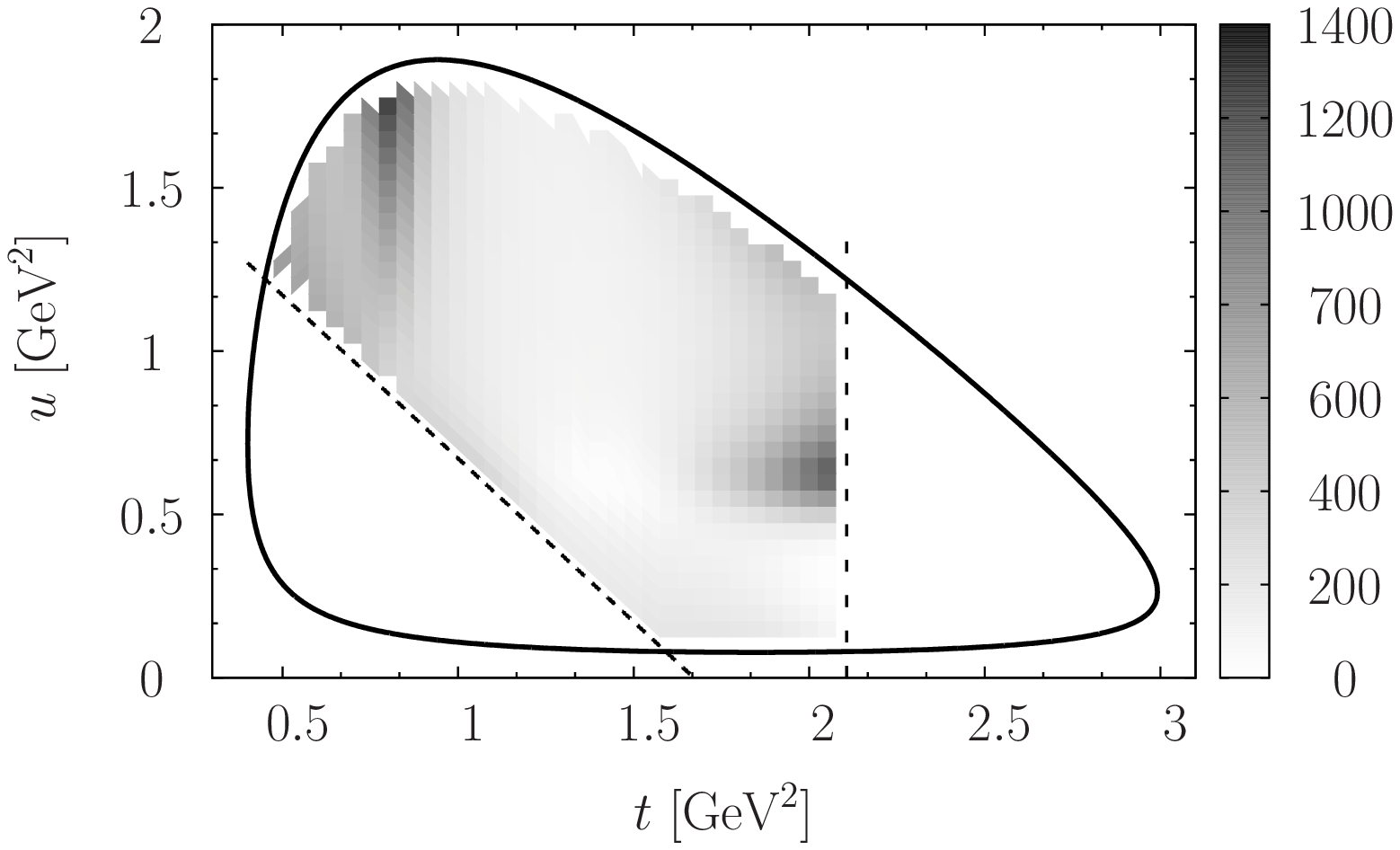}
\caption{The experimental (upper plot) and theoretical Dalitz plot fitted to the data (fit~2) (lower plot) are shown. 
The dashed lines denote the restriction of the fits to the region $(s,t)<(M_{\eta'} +\mK)^2$. }
\label{fig:Dalitz_BES}
\end{figure}
\begin{table}[t]
\centering
\renewcommand{\arraystretch}{1.4}
 \begin{tabular}{c c c }
 \toprule
 		       & Fit 1				& Fit 2  \\			
 \midrule
 $|c_0|\times\text{GeV}^2$	       &$0.17\pm0.01$  		&$0.21\pm0.08$		\\
 $|c_1|\times\text{GeV}^4$      &$0.26\pm0.03 $		 	&$0.21\pm0.09 $ \\
 $c_2$ 		       	     	& 1 (fixed) 			& 1 (fixed)\\
 $|c_3|\times\text{GeV}^2$      &$1.80\pm0.05 $			&$1.63\pm0.08 $ \\
 $|c_4|\times\text{GeV}^4$      &$0.88\pm0.08 $    		&$0.76\pm0.12 $ \\
 $|c_5|\times\text{GeV}^6$      &$0.20\pm0.05 $	     	 	&$0.15\pm0.05 $	\\ 
 $|c_6|\times10^2\text{GeV}^4$  &$3\pm6 $			&$5\pm2 $	\\
  arg\,$c_0$		        &$0.4\pm0.1 $ 			&$0.5\pm0.2 $	\\
  arg\,$c_1$		        &$-0.5\pm0.3 $  		&$-0.2\pm0.3 $	  \\
  arg\,$c_3$		        &$0.07\pm0.02 $ 		&$0.05\pm0.01 $\\ 
  arg\,$c_4$		        &$0.03\pm0.01 $			&$0.02\pm0.05 $ \\  
  arg\,$c_5$			&$-0.12\pm0.04 $		&$-0.14\pm0.21$ \\ 
  arg\,$c_6$		        &$-0.05\pm0.02 $		&$-0.14\pm0.46 $ \\ 
 \midrule
 $\chi^2/\text{d.o.f.}	$  	&$1.27\pm0.01 $ 		&$1.35\pm0.07 $	\\  
 
 \bottomrule
 \end{tabular}
\renewcommand{\arraystretch}{1.0}
\caption{\textit{Fit to BESIII data:} Numerical fit results for the subtraction constants $c_i$  and the corresponding $\chi^2/\text{d.o.f.}$. Two fit scenarios are considered: dispersive fits without $D$-wave (fit~1) and with $D$-wave (fit~2).
The uncertainties of the input scattering phases are included in the resulting subtraction constant errors.}
\label{tab:tabFITresultBES}
\end{table}

\begin{table}[t]
\centering
\renewcommand{\arraystretch}{1.4}
\begin{tabular}{c c c }
\toprule
  		&Fit 1 		&Fit 2 		\\  
\midrule
$\FF_0^2$	&$(5\pm 2) \%$		&$(4.6\pm 0.3)\%$ \\				
$\FF_1^1$ 	&$(21\pm  5) \%$	&$(16\pm 3) \%$	 \\					
$ \FF_0^{1/2}$	&  $(39\pm 5)\%$ & $(43\pm4)\%$ \\
$\FF_1^{1/2}$	&  $(9.2\pm 0.5)\%$ & $(7\pm 2)\%$ \\
$\FF_0^{3/2}$ 	&  $(6\pm 2)\%$  &  $(9\pm 3)\%$ \\
$ \FF_2^{1/2}$	& ---  & $(1.52\pm0.05)\% $ \\
\bottomrule
\end{tabular}
\renewcommand{\arraystretch}{1.0}
\caption{\textit{Fit fractions BESIII:} The resulting fit fractions for the different fit scenarios. The uncertainties of the input scattering phases are included in the resulting subtraction constant errors.
The fit fractions of the $\pi K$ amplitudes in the $s$- and $t$-channel are summed together.}
\label{tab:tabFitfrac_BES}
\end{table}

We begin with the comparison to the Dalitz plot data of the $D^+ \to K^0_S\pi^0\pi^+$ decay measured by the BESIII collaboration~\cite{Ablikim:2014cea}.
The experimental Dalitz plot, shown in Fig.~\ref{fig:Dalitz_BES} (upper plot), exhibits two prominent resonant structures, the $\rho(770)$ and $K^*(892)$ resonances.
Compared to the $D^+\to K^-\pi^+\pi^+$ Dalitz plot, which shows only a prominent $K^*(892)$ resonance, we have direct access to the $\pi\pi$ $P$-wave and therefore a much stronger constraint on 
the subtraction constants $c_0$ and $c_1$. 
Since our dispersive approach requires elastic two-particle unitarity, Eq.~\eqref{eq:unrel}, we restrict the fit region to pion--kaon two-particle energies below the $\eta' K$ threshold, above which
the major onset of inelasticities is seen phenomenologically~\cite{Jamin:2001zq,Edera:2005dk,Moussallam:2007qc}, in particular in the $S$-wave.
We can therefore assume elastic unitarity to be a good approximation below this threshold.
The provided data set comprises a binned $t\times u$ Dalitz plot with bin size $0.05\,\text{GeV}^2\times 0.05\,\text{GeV}^2$.
We define the event distribution function of the efficiency- and background-corrected data by
\begin{align}\label{eq:PDFBES}
 \mathcal{P}(t_i,u_i)=\int_{t_i-\delta}^{t_i+\delta} \int_{u_i-\delta}^{u_i+\delta}
|\M_{\bar{0}0+}(s(t,u),t,u)|^2 \diff u\,\diff t\,,
\end{align}
with $(t_i,u_i)$ being the center of the corresponding bin and $2\delta=0.05\,\text{GeV}^2$ the bin width.

We perform  two fit scenarios: without the $\pi K$ $D$-wave $\F^{1/2}_2(x)$ (fit~1) and including the $D$-wave (fit~2), analogously to Ref.~\cite{Niecknig:2015ija}.
The fit region is confined to $t,s<(M_{\eta'} +\mK)^2$ and the $\chi^2$ given by
\beq
 \chi^2=\sum_{i=1}^{746} \frac{\big[\N\mathcal{P}(t_i,u_i)-(\text{\#corrected events/bin})_i\big]^2}{(\text{\#corrected events/bin})_i}\,,
\eeq
where $\N$ is the overall normalization of the corrected data with 746 bins in the considered fit region.
Furthermore we define fit fractions as follows:
\beq \label{eq:fitfrac}
\FF^I_J=\frac{\int|P_J(x(s,t))\,\F^I_{J}(x(s,t))|^2\, \diff s\,\diff t}{\int |\M_{\bar{0}0+}(s,t,u)|^2\, \diff s\,\diff t}\,, 
\eeq
where the $P_J(x)$ denote the angle-dependent prefactors of the corresponding single-variable amplitudes in the total amplitude.

The fit results for the subtraction constants are summarized in Table~\ref{tab:tabFITresultBES}, the emanating fit fractions are displayed in Table~\ref{tab:tabFitfrac_BES}. In Fig.~\ref{fig:Dalitz_BES} (lower plot) we display the fitted theoretical Dalitz plot.
The two fits show only little difference in the values of the subtraction constants and fit fractions, in contrast to our findings in the earlier $D^+\to K^-\pi^+\pi^+$ analysis~\cite{Niecknig:2015ija}, where the $D$-wave had a sizable 
impact on both subtraction constants and fit fractions. Including the $D$-wave actually worsens the fit quality from $\chi^2/\text{d.o.f.}=1.27\pm0.01$ (without $D$-wave) to $\chi^2/\text{d.o.f.}=1.35\pm0.07$ (with $D$-wave); remember that the $D$-wave does not include any additional free parameter in the amplitude representation~\eqref{eq:fulleq}.
However, no region of particular disagreement is observed in the Dalitz plot.
The moduli of the subtraction constants differ significantly from the ones extracted from $D^+\to K^-\pi^+\pi^+$~\cite{Niecknig:2015ija} and show smaller uncertainties.
However, the phases of the subtraction constants attained in the \mbox{BESIII} fit are compatible (modulo $2\pi$) with the CLEO/FOCUS phases of Ref.~\cite{Niecknig:2015ija}, with the exception of the phase of $c_1$.
We note that $c_1$ is the linear $\pi\pi$ $P$-wave subtraction constant, which contributes only indirectly via (charge-exchange) rescattering to the $D^+\to K^-\pi^+\pi^+$  decay amplitude.
The phases of the $\F_0^{1/2}$ subtraction constants mutually agree modulo $\pi$ and can be chosen nearly real with an overall phase factored out, similar to what has been observed in Ref.~\cite{Niecknig:2015ija}. 
However, this does not hold for the $\F_1^1$ amplitude.

Strictly speaking, due to isospin symmetry, the CLEO, FOCUS, and BESIII fits should result in the same values for the subtraction constants. 
Seeing that the BESIII fit clashes with the combined CLEO/FOCUS results of Ref.~\cite{Niecknig:2015ija},
it is however doubtful that a combined fit proves to be successful.
With the overall $\chi^2$ given by the sum of the individual $\chi^2$ values,
we attempt to perform a simultaneous fit of all three data sets (CLEO, FOCUS, and BESIII) available to us.
The fit results in $\chi^2_\text{combined}/\text{d.o.f.}$ values of $1.7\pm0.1$ ($2.5\pm0.2$) for fit~1 (fit~2): the inclusion of the $\pi K$ $D$-wave in the combined fit considerably worsens the quality of the data description. 
This suggests that the heuristic inclusion of the $\pi K$ $D$-wave,
which is necessary to obtain sensible fit fractions in the CLEO fit~\cite{Niecknig:2015ija}, 
may seem sufficient for the individual fits, but is clearly not for a combined analysis.

\section{Alternative \boldmath{$D$}-wave model}\label{sec:DwaveOm}

\begin{table}[t]
\centering
\renewcommand{\arraystretch}{1.4}
\begin{tabular}{c c c  }
\toprule
			       & BESIII 				               		&  combined fit		\\  
\midrule
$|c_0|\times\text{GeV}^2$	&$0.17\pm0.02$  				                       &$0.15\pm0.02$ 	             		\\  
$|c_1|\times\text{GeV}^4$      	&$0.19\pm0.04$		 	 		                       &$0.21\pm0.03$					\\  
$c_2$				& 1 (fixed) 				                       & 1 (fixed) 		      		\\  
$|c_3|\times\text{GeV}^2$      	&$1.61\pm0.06$					     			&$1.78\pm0.03$			\\  
$|c_4|\times\text{GeV}^4$      	&$0.74\pm0.08$    			 		        &$0.89\pm0.04$       \\  
$|c_5|\times\text{GeV}^6$      	&$0.18\pm0.02$	     	 					&$0.19\pm0.02$		\\ 
$|c_6|\times10^2\text{GeV}^4$  	&$3.4\pm 0.3$						  	&$3\pm 1$	\\ 
$|c_7|\times10^3\text{GeV}^8$ 	&$9\pm 4$							&$9\pm3$		\\  
arg\,$c_0$		        &$0.33\pm0.07$ 				 			&$0.33\pm0.12  $	\\ 
arg\,$c_1$		        &$-0.21\pm0.17$				&$-0.36\pm0.11$		\\  
arg\,$c_3$		        &$-0.07\pm0.02$ 		  				&$-0.10\pm0.01$		\\  
arg\,$c_4$		        &$0.02\pm0.02$				&$-0.11\pm0.01$	\\  
arg\,$c_5$		        &$-0.07\pm0.06$							&$-0.10\pm0.02$		\\  
arg\,$c_6$		        &$-0.7\pm0.3$							&$-0.0\pm0.3$		\\  
 arg\,$c_7$                     &$1.3\pm0.3$							&$-1.4\pm0.2$		\\  
\midrule
$\chi_\mathrm{CLEO}^2/\text{d.o.f.}$  	&--- 							&$1.19\pm0.03$	\\  
$\chi_\mathrm{FOCUS}^2/\text{d.o.f.}$	&--- 							&$1.28\pm0.01$ \\
$\chi_\mathrm{BES}^2/\text{d.o.f.}$	&$1.08 \pm 0.01$ 					&$1.26\pm0.04$ \\
\bottomrule
$\chi_\mathrm{combined}^2/\text{d.o.f.}$	&$1.08 \pm0.01$						&$1.22\pm0.03$\\
\bottomrule
\end{tabular}
\renewcommand{\arraystretch}{1.0}
\caption{\textit{Alternative $D$-wave fits:} Numerical fit results for the subtraction constants $c_i$ and the corresponding individual and combined $\chi^2/\text{d.o.f.}$. Two fit scenarios are considered: fit to the BESIII data only 
and combined fit to the CLEO, FOCUS, and BESIII data sets. The errors on the input scattering phase shifts are again included in the subtraction constant errors.}
\label{tab:tabFITDwave}
\end{table}

\begin{table}
 \centering
 \renewcommand{\arraystretch}{1.4}
   \setlength{\tabcolsep}{0.9mm}
 \begin{tabular}{c c c c}
 \toprule
  {} 		&{BESIII$_\text{individual}$}&{BESIII$_\text{combined}$} &	 {CLEO/FOCUS$_\text{combined}$}	\\  
 \midrule
 $\FF_0^2$		&$(4\pm 2)\%$		&$(1.7\pm 0.5)\%$  	& $(5\pm 1)\%$ 		\\
 $\FF_1^1$		&$(17\pm 2)\%$		&$(23\pm 3)\%$		&--- 			\\
 $\FF_0^{1/2}$		&$(48\pm 2)\%$		&$(36\pm 5)\%$		&$(46\pm 6)\%$		\\
 $\FF_1^{1/2}$		&$(7.5\pm 0.5)\%$	&$(8.5\pm 0.4)\%$	&$(11.5\pm 0.5)\%$	\\
 $\FF_0^{3/2}$ 		&$(10\pm 1)\%$		&$(6\pm 1)\%$		&$(0.6\pm 0.1)\%$	\\
 $\FF_2^{1/2}$ 		&$(0.4\pm 0.1)\%$	&$(0.5\pm 0.1)\%$	&$(0.7\pm 0.1)\%$	\\
 \bottomrule
 \end{tabular}
 \renewcommand{\arraystretch}{1.0}
 \caption{\textit{Alternative $D$-wave fit fractions:} The resulting fit fractions for the different fit scenarios: individual fits to the BESIII data (left column) and combined fit to all three data sets simultaneously (middle and right column). The errors on the parameters are evaluated by varying the basis functions within their error bands.
 The fit fractions of the $\pi K$ amplitudes in the $s$- and $t$-channel are summed together.}
 \label{tab:tabFitfrac_Dwave}
\end{table}

In this section we want to assess the origin of the bad fit qualities in the combined analysis. Since the prime candidate is the only partially included $\pi K$ $I=1/2$ $D$-wave, we attempt to improve on the latter by oversubtracting
the amplitude $\F_2^{1/2}$ once, with the aim to obtain a more flexible description of the $D$-wave strength:
\begin{equation}     
 \F_2^{1/2}(s)=\Omega^{1/2}_2(s)\Bigg\{c_7+\frac{s}{\pi}\int_{s_\mathrm{th}}^\infty \frac{\diff s'}{s'}
 \frac{\hat{\F}_2^{1/2}(s')\sin\delta^{1/2}_2(s')}{\big|\Omega^{1/2}_2(s')\big|(s'-s)}\Bigg\}\,.\label{eq:F2sub}
\end{equation}
Essentially this is equivalent to adding the $\pi K$ $D$-wave Omn\`es function times a (complex) normalization constant $c_7$ to the original set of amplitudes.  In principle, this worsens the high-energy behavior of the $D$-wave contribution even more; in practice, given our prescription to drop the $D$-wave contributions to the inhomogeneities, it does not render this inconsistency any more severe.

We have previously justified the inclusion of the $D$-wave amplitude as in Eq.~\eqref{eq:fulleq}---retaining an inhomogeneity generated by the crossed-channel $S$- and $P$-waves, while neglecting the projections of $\F_2^{1/2}$ onto the other partial waves---by alluding to low-energy processes such as those calculated in chiral perturbation theory~\cite{Niecknig:2015ija}. For instance, the $I=0$ $\pi\pi$ $D$-wave scattering length is almost completely given by crossed-channel dynamics~\cite{Gasser:1983yg,Colangelo:2001df}, not by the low-energy tail of the $f_2(1270)$ resonance.  Similarly, but outside the realm of applicability of chiral dynamics, the subleading $F$-wave in $\omega\to 3\pi$ is dominated by crossed-channel amplitudes, not by the $\rho_3(1690)$~\cite{Niecknig:2012sj}. To assume that this picture could be extended beyond the near-threshold region, up to energies where the $D$-wave becomes resonant, was clearly too optimistic. In this sense, the new subtraction constant $c_7$ could be linked to an independent coupling constant to the $K_2(1430)$ tensor resonance: it is the only parameter that allows to adjust the strength of the $\pi K$ $D$-wave independently of the crossed-channel dynamics.

We again consider two fit scenarios: a fit to the BESIII data alone and a combined fit to the CLEO, FOCUS, and BESIII data sets.
The results are summarized in Table~\ref{tab:tabFITDwave}, and the ensuing fit fractions in Table~\ref{tab:tabFitfrac_Dwave}.

The BESIII fit results for the subtraction constants turn out to be very similar to the BESIII fits~1 and 2, see Table~\ref{tab:tabFITresultBES} for comparison. 
This is an anticipated result, since in the previous fit scenarios (fits~1 and 2) the inclusion of the $D$-wave did not change the subtraction constants substantially,
but led to a poorer $\chi^2$ result. 
Similarly the fit fractions, comparing the previous fit scenario~2 and the alternative $D$-wave BESIII fit, are alike, with the exception of the $\F_2^{1/2}$ single-variable amplitude, which is slightly smaller.
The fit quality with the more flexible $D$-wave is improved to $\chi_\mathrm{BES}^2/\text{d.o.f.}=1.08\pm0.01$ compared to $\chi^2/\text{d.o.f.}=1.27\pm0.01$ ($\chi^2/\text{d.o.f.}=1.35\pm0.07$) for fit~1 (fit~2) in the previous section.

With all data sets combined, we obtain a combined $\chi^2/\text{d.o.f.}=1.22\pm0.03$, which presents a huge improvement over the previous combined fits~1/2  ($\chi^2/\text{d.o.f.}=1.7\pm0.1 / 2.5\pm0.2$).
The $\chi^2$ thus advocates that the discrepancy in fits~1/2 comes from the heuristically built-in $D$-wave.
However, we note again that a thorough inclusion of the $D$-wave in the Khuri--Treiman formalism would necessitate further subtraction constants in the other partial waves.

The subtraction-constant results are very similar to the individual BESIII fit values, but the  $\chi_\mathrm{BES}^2$  worsens from $1.08\pm0.01$ to $1.26\pm0.04$.
In contrast, we find that the obtained  $\chi_\mathrm{CLEO}^2$ and $\chi_\mathrm{FOCUS}^2$ values are similar to the individual fit qualities of Ref.~\cite{Niecknig:2015ija}, although the subtraction constants are significantly different.
This suggests that the $D^+\to \bar{K}^0\pi^0\pi^+$ data constrains the subtraction constants much better than the available $D^+\to K^-\pi^+\pi^+$ one.
The fit fractions for the $D^+\to \bar{K}^0\pi^0\pi^+$ decay amplitude are very similar to the BESIII fit while the large constructive interference effects seen in the CLEO/FOCUS fits of Ref.~\cite{Niecknig:2015ija} do not show up as prominently
in the $D^+\to K^-\pi^+\pi^+$  fit fractions.	
Additionally, we observe that the fit fractions of the non-resonant waves $\F_0^2$ and $\F_0^{3/2}$ reduce compared to the individual BESIII and CLEO/FOCUS fits.

Clearly, our $\chi^2/\text{d.o.f.}=1.22\pm0.03$ for the combined fit, though largely improved, is still not remarkably good, given the large number of data points. To put this into perspective, however, we wish to point out that most previous theoretical studies refrained from fitting Dalitz plots at all~\cite{Oller:2004xm,Boito:2009qd,Magalhaes:2011sh,Magalhaes:2015fva,Guimaraes:2014cca}, while Ref.~\cite{Nakamura:2015qga} performs a fit to pseudodata that does not allow for a sensible statistical interpretation. Isobar-model fits to the FOCUS~\cite{FOCUS,Link:2009ng} and BESIII~\cite{Ablikim:2014cea} data (individually) by the experimental collaborations result in $\chi_\mathrm{FOCUS}^2/\text{d.o.f.} = 1.17$ and $\chi_\mathrm{BES}^2/\text{d.o.f.} = 1.41$, respectively, and are therefore hardly better or worse than what we find.

\begin{figure*}
 \centering
 \includegraphics*[width=0.925\linewidth]{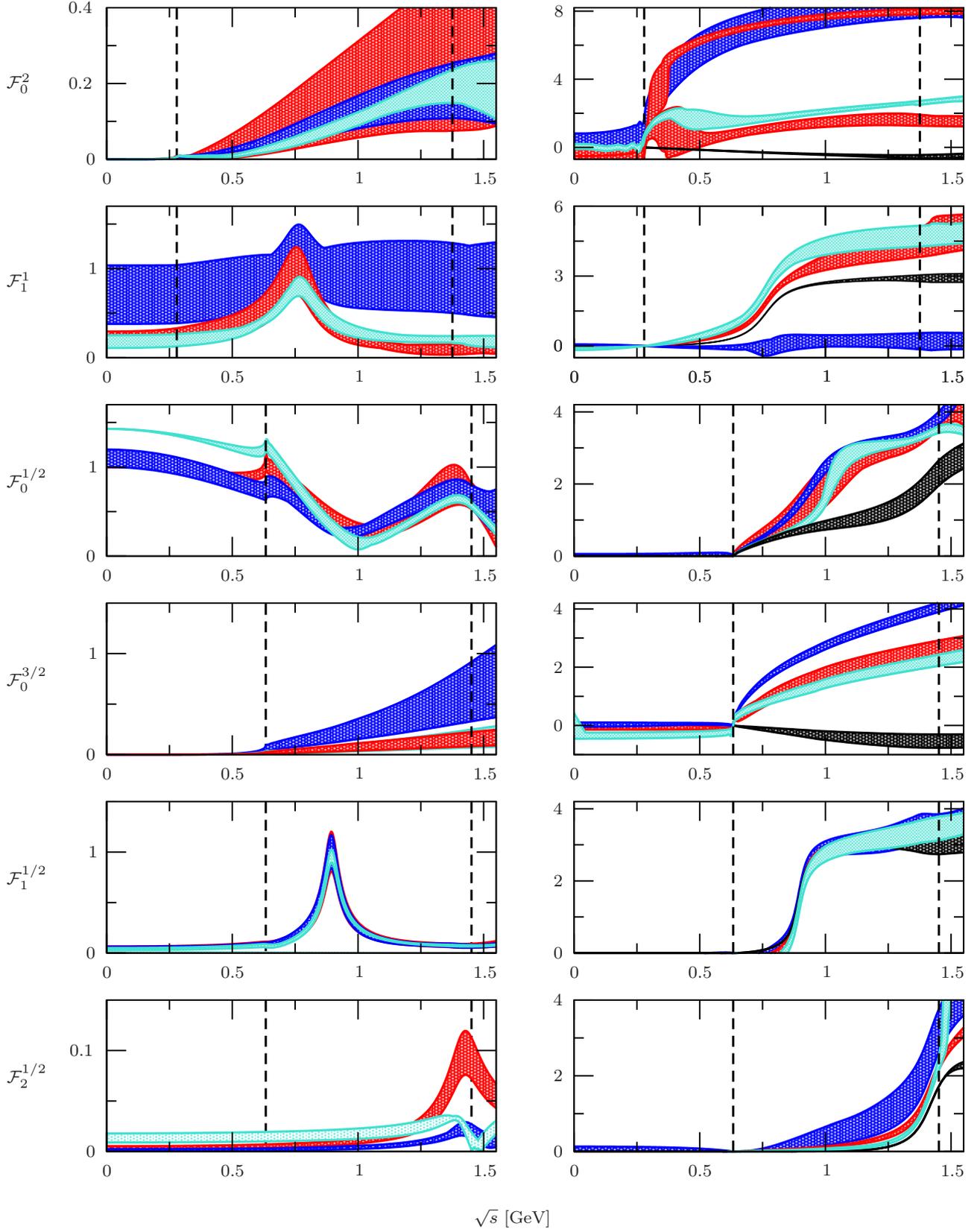}
 \caption{\textit{Left column:} Absolute values of the single-variable amplitudes of fit~2,  in arbitrary units: BESIII in red, combined fit with alternative $D$-wave turquoise, and the CLEO/FOCUS fits of Ref.~\cite{Niecknig:2015ija} in blue. The overall normalization is chosen such that the absolute values in the $K^*(892)$ peak in the $I=1/2$ $P$-wave agree.
 \textit{Right column:} Phase motion of the single-variable amplitudes (BESIII in red, alternative $D$-wave turquoise, CLEO/FOCUS in blue) and input scattering phases (black) in radiant. The phases are fixed to zero at the two-particle ($\pi\pi$, $\pi K$) thresholds. Note that we obtain two separate solutions for the $\F_0^2$ phase.
 The dashed lines visualize the fitted area; for the $\pi K$ amplitudes from threshold to the $\eta' K$ threshold  and the full phase space for the $\pi\pi$ amplitudes.}
 \label{fig:PWBES}
\end{figure*}

In Fig.~\ref{fig:PWBES}, we show moduli and phases of the extracted single-variable amplitudes, comparing the results of the CLEO/FOCUS
fits of Ref.~\cite{Niecknig:2015ija} to fit~2 to the BESIII data as well as the combined fit to all three data sets including
the alternative $D$-wave description.  For the purpose of comparison, the overall normalizations of the total amplitudes are chosen such that in Fig.~\ref{fig:PWBES}, the peak strengths
of the $K^*(892)$ resonance in the $I=1/2$ $P$-wave coincide in all fits.  The main observation is that the new fits of the present study
including the BESIII data, while consistent with the previous results, constrain most amplitudes much better; 
above all others, this rather obviously holds true for the $\pi\pi$ $P$-wave.  
An important result is that the $D^+\to K_S \pi^0\pi^+$ Dalitz plot confirms the conclusion of Ref.~\cite{Niecknig:2015ija} concerning
the phase motion of the $I=1/2$ $\pi K$ $S$-wave, whose phase rises much more quickly than the elastic scattering phase 
shift~\cite{Aston:1987ir}.  This has been observed before~\cite{Aitala:2005yh,Link:2009ng}, but does not contradict Watson's final-state theorem~\cite{Watson:1954uc}: the phase is modified due to rescattering with the third particle in the final state.  

The $\pi K$ $D$-wave merits a further comment, as the different fits in Fig.~\ref{fig:PWBES} display quite
different magnitudes of $\F_2^{1/2}$: in particular the fit to the BESIII data alone yields a much larger contribution
than the others.
Remember that before the oversubtraction of Eq.~\eqref{eq:F2sub}, this is indirectly determined by the other 
partial waves.  While the $D^+\to K_S \pi^0\pi^+$ data in the BESIII fit show rather little sensitivity to the $D$-wave,
but strong constraints on the subtraction constants, in particular the fit fractions in the $D^+\to K^- \pi^+\pi^+$ 
data depend significantly thereon~\cite{Niecknig:2015ija}.  Only the more flexible form allows to reconcile both, with destructive 
interference between the $D$-wave generated through crossed-channel rescattering and the new subtraction constant
$c_7$.

Finally, the seemingly stark modifications of the phases of the small, nonresonant amplitudes must not be overstated: they occur in places where the moduli are close to zero, hence the full amplitudes are not severely affected.

\section{Conclusion}\label{sec:conclusion}
In this letter we have resumed the study of strong final-state interactions in the  $D^+\to K^- \pi^+ \pi^+/\bar{K}^0 \pi^0 \pi^+$ decays 
utilizing  dispersion relations in the form of Khuri--Treiman-type equations.
The resulting dispersion relations satisfy analyticity, unitarity, and crossing symmetry by construction and generate crossed-channel rescattering between the three final-state particles.
We solely rely on $\pi\pi$ and $\pi K$ scattering phase-shift input.  The seven complex subtraction constants
are fitted to the $D^+\to K^- \pi^+ \pi^+/\bar{K}^0 \pi^0 \pi^+$ data from the CLEO, FOCUS, and \mbox{BESIII} collaborations.
The requirement of elastic unitarity restricts the comparison to energies below the onset of major inelasticities.
We have focused on two main fit scenarios: the individual fit to the  $D^+\to \bar{K}^0 \pi^0 \pi^+$ Dalitz plot from the \mbox{BESIII} collaboration, and a combined fit to all the 
data sets available to us. In both scenarios we have in particular studied the impact of the $\pi K$ $D$-wave. 

The individual fits to the BESIII $D^+\to \bar{K}^0 \pi^0 \pi^+$ data show that the Dalitz plot in the elastic region can be described reasonably well with an overall $\chi^2/\text{d.o.f.}$ of around $1.3$. 
The subtraction constants are much more constrained than previously by $D^+\to K^- \pi^+ \pi^+$ data alone,  
and much more stable with respect to inclusion of the $\pi K$ $D$-wave, which does not improve the fit quality.
In particular the direct sensitivity to the $\pi\pi$ $P$-wave has a large impact.
We confirm once more that crossed-channel rescattering effects significantly shape the phases of the partial wave amplitudes;
in particular the isospin $1/2$ $\pi K$ $S$-wave phase shows a much steeper rise compared to the elastic scattering phase shift.

We find, however, that the values of the subtraction constants do not agree between the two differently charged final states, 
and a combined fit does not lead to satisfactory results.
This tension could be traced back to the $\pi K$ D-wave, which is not fully consistently included in our Khuri--Treiman system.
Introducing an additional subtraction in the $\pi K$ $D$-wave, hence allowing for an additional free parameter therein, 
we demonstrate that the description of the BESIII data improves to $\chi^2/\text{d.o.f.} \approx 1.1$ in the individual fit. 
Consistency of all three data sets (and both charge channels) is reinstated, 
with an overall $\chi^2/\text{d.o.f.}$ of about $1.2$.

We have therefore demonstrated successful steps towards the application of Khuri--Treiman-type amplitude representations to three-body 
decays of charmed mesons, exploiting the consequences of isospin symmetry and the coupling of the various partial wave
through rescattering in an optimal way.  Further, systematic applications of this formalism should be investigated in the future. These should expand on the effects of inelasticities and coupled channels~\cite{Albaladejo:2017hhj,Guo:2015kla} in order to extend the dispersive description to the complete $D$-meson Dalitz plots, investigate the effects of three-body unitarity, and search for ways to include higher partial waves consistently without a proliferation of free parameters.

\section*{Acknowledgments}

We thank the FOCUS, CLEO, and BESIII collaborations for providing us with the Dalitz
plot data of Refs.~\cite{FOCUS,CLEO,Ablikim:2014cea}, respectively,
as well as Chengdong Fu and Wolfgang Gradl for helpful e-mail communication concerning the data of 
Ref.~\cite{Ablikim:2014cea}.
Financial support by DFG and NSFC through funds provided to the Sino--German CRC~110 ``Symmetries
and the Emergence of Structure in QCD'' (DFG Grant
No.~TRR110 and NSFC Grant No.~11621131001) is gratefully acknowledged.


\begin{thebibliography}{99}
\biboptions{sort&compress}

\bibitem{Battaglieri:2014gca}
  M.~Battaglieri \textit{et al.},
  Acta Phys.\ Polon.\ B \textbf{46} (2015) 257
  [arXiv:1412.6393 [hep-ph]].

\bibitem{Caprini:2005zr}
  I.~Caprini, G.~Colangelo and H.~Leutwyler,
  Phys.\ Rev.\ Lett.\  \textbf{96} (2006) 132001
  [hep-ph/0512364].

\bibitem{GarciaMartin:2011jx}
  R.~Garc\'ia-Mart\'in, R.~Kami\'nski, J.~R.~Pel\'aez, and J.~Ruiz de Elvira,
  Phys.\ Rev.\ Lett.\  \textbf{107} (2011) 072001
  [arXiv:1107.1635 [hep-ph]].

\bibitem{DescotesGenon:2006uk}
  S.~Descotes-Genon and B.~Moussallam,
  Eur.\ Phys.\ J.\ C \textbf{48} (2006) 553
  [hep-ph/0607133].

\bibitem{Pelaez:2016klv}
  J.~R.~Pel\'aez, A.~Rodas and J.~Ruiz de Elvira,
  Eur.\ Phys.\ J.\ C \textbf{77} (2017) 91
  [arXiv:1612.07966 [hep-ph]].

\bibitem{GardnerMeissner}
  S.~Gardner and U.-G.~Mei{\ss}ner,
  Phys.\ Rev.\  D \textbf{65} (2002) 094004
  [arXiv:hep-ph/0112281].

\bibitem{Daub:2015xja}
  J.~T.~Daub, C.~Hanhart and B.~Kubis,
  JHEP \textbf{1602} (2016) 009
  [arXiv:1508.06841 [hep-ph]].
 
\bibitem{Albaladejo:2016mad}
  M.~Albaladejo, J.~T.~Daub, C.~Hanhart, B.~Kubis and B.~Moussallam,
  JHEP \textbf{1704} (2017) 010
  [arXiv:1611.03502 [hep-ph]].

\bibitem{KhuriTreiman}
  N.~N.~Khuri and S.~B.~Treiman,
  Phys.\ Rev.\  \textbf{119} (1960) 1115.
  
\bibitem{Bronzan} 
  J.~B.~Bronzan and C.~Kacser,
  Phys.\ Rev.\ \textbf{132} (1963) 2703.
  
\bibitem{Aitchison:1965zz}
  I.~J.~R.~Aitchison,
  Phys.\ Rev.\  \textbf{137} (1965) B1070.

\bibitem{Aitchison:1966lpz}
  I.~J.~R.~Aitchison and R.~Pasquier,
  Phys.\ Rev.\ \textbf{152} (1966) 1274.

\bibitem{Pasquier:1968zz}
  R.~Pasquier and J.~Y.~Pasquier,
  Phys.\ Rev.\  \textbf{170} (1968) 1294.
  
\bibitem{Pasquier:1969dt}
  R.~Pasquier and J.~Y.~Pasquier,
  Phys.\ Rev.\  \textbf{177} (1969) 2482.

\bibitem{Aitchison:2015jxa}
  I.~J.~R.~Aitchison,
  arXiv:1507.02697 [hep-ph].

\bibitem{AnisovichLeutwyler}
  A.~V.~Anisovich and H.~Leutwyler,
  Phys.\ Lett.\ B \textbf{375} (1996) 335
  [hep-ph/9601237].

\bibitem{Kambor:1995yc}
  J.~Kambor, C.~Wiesendanger and D.~Wyler,
  Nucl.\ Phys.\ B \textbf{465} (1996) 215
  [hep-ph/9509374].

\bibitem{Guo:2015zqa}
  P.~Guo, I.~V.~Danilkin, D.~Schott, C.~Fern\'andez-Ram\'irez, V.~Mathieu and A.~P.~Szczepaniak,
  Phys.\ Rev.\ D \textbf{92} (2015) 054016
  [arXiv:1505.01715 [hep-ph]].

\bibitem{Guo:2016wsi}
  P.~Guo, I.~V.~Danilkin, C.~Fern\'andez-Ram\'irez, V.~Mathieu and A.~P.~Szczepaniak,
  Phys.\ Lett.\ B {\bf 771} (2017) 497
  [arXiv:1608.01447 [hep-ph]].

\bibitem{Colangelo:2016jmc}
  G.~Colangelo, S.~Lanz, H.~Leutwyler and E.~Passemar,
  Phys.\ Rev.\ Lett.\  \textbf{118} (2017) 022001
  [arXiv:1610.03494 [hep-ph]].

\bibitem{Albaladejo:2017hhj}
  M.~Albaladejo and B.~Moussallam,
  Eur.\ Phys.\ J.\ C {\bf 77} (2017) 508
  [arXiv:1702.04931 [hep-ph]].

\bibitem{Niecknig:2012sj}
  F.~Niecknig, B.~Kubis and S.~P.~Schneider,
  Eur.\ Phys.\ J.\ C \textbf{72} (2012) 2014
  [arXiv:1203.2501 [hep-ph]].
  
\bibitem{Danilkin:2014cra}
  I.~V.~Danilkin, C.~Fern\'andez-Ram\'irez, P.~Guo, V.~Mathieu, D.~Schott, M.~Shi and A.~P.~Szczepaniak,
  Phys.\ Rev.\ D \textbf{91} (2015) 094029
  [arXiv:1409.7708 [hep-ph]].

\bibitem{Schneider:2012nng}
  S.~P.~Schneider, 
  PhD thesis, University of Bonn (2013)
  [http://hss.ulb.uni-bonn.de/2013/3126/3126.htm].

\bibitem{etaprime}
  T.~Isken, B.~Kubis, S.~P.~Schneider and P.~Stoffer,
  Eur.\ Phys.\ J.\ C \textbf{77} (2017) 489
  [arXiv:1705.04339 [hep-ph]].

\bibitem{Niecknig:2015ija}
  F.~Niecknig and B.~Kubis,
  JHEP \textbf{1510} (2015) 142
  [arXiv:1509.03188 [hep-ph]].

\bibitem{Oller:2004xm}
  J.~A.~Oller,
  Phys.\ Rev.\ D \textbf{71} (2005) 054030
  [hep-ph/0411105].

\bibitem{Boito:2009qd}
  D.~R.~Boito and R.~Escribano,
  Phys.\ Rev.\ D \textbf{80} (2009) 054007
  [arXiv:0907.0189 [hep-ph]].
  
\bibitem{Magalhaes:2011sh}
  P.~C.~Magalh\~{a}es, M.~R.~Robilotta, K.~S.~F.~F.~Guimar\~{a}es, T.~Frederico, W.~de Paula, I.~Bediaga, A.~C.~dos Reis and C.~M.~Maekawa, 
  Phys.\ Rev.\ D \textbf{84} (2011) 094001
  [arXiv:1105.5120 [hep-ph]].
  
\bibitem{Magalhaes:2015fva}
  P.~C.~Magalh\~{a}es and M.~R.~Robilotta,
  Phys.\ Rev.\ D \textbf{92} (2015) 094005
  [arXiv:1504.06346 [hep-ph]].
  
\bibitem{Guimaraes:2014cca}
  K.~S.~F.~F.~Guimar\~{a}es, O.~Louren\c{c}o, W.~de Paula, T.~Frederico and A.~C.~dos Reis,
  JHEP \textbf{1408} (2014) 135
  [arXiv:1404.3797 [hep-ph]].

\bibitem{ElBennich:2006yi}
  B.~El-Bennich, A.~Furman, R.~Kami\'nski, L.~Le\'sniak and B.~Loiseau,
  Phys.\ Rev.\ D \textbf{74} (2006) 114009
  [hep-ph/0608205].

\bibitem{ElBennich:2009da}
  B.~El-Bennich, A.~Furman, R.~Kami\'nski, L.~Le\'sniak, B.~Loiseau and B.~Moussallam,
  Phys.\ Rev.\ D \textbf{79} (2009) 094005
   [Phys.\ Rev.\ D \textbf{83} (2011) 039903]
  [arXiv:0902.3645 [hep-ph]].

\bibitem{Aitala:2005yh}
  E.~M.~Aitala \textit{et al.}  [E791 Collaboration],
  Phys.\ Rev.\ D \textbf{73} (2006) 032004
   [Phys.\ Rev.\ D \textbf{74} (2006) 059901]
  [hep-ex/0507099].

\bibitem{FOCUS}
  J.~M.~Link \textit{et al.}  [FOCUS Collaboration],
  Phys.\ Lett.\ B \textbf{653} (2007) 1
  [arXiv:0705.2248 [hep-ex]].

\bibitem{CLEO}
  G.~Bonvicini \textit{et al.}  [CLEO Collaboration],
  Phys.\ Rev.\ D \textbf{78} (2008) 052001
  [arXiv:0802.4214 [hep-ex]].
   
\bibitem{Link:2009ng}
  J.~M.~Link \textit{et al.} [FOCUS Collaboration],
  Phys.\ Lett.\ B \textbf{681} (2009) 14
  [arXiv:0905.4846 [hep-ex]].

\bibitem{Nakamura:2015qga}
  S.~X.~Nakamura,
  Phys.\ Rev.\ D \textbf{93} (2016) 014005
  [arXiv:1504.02557 [hep-ph]].

\bibitem{Ablikim:2014cea}
  M.~Ablikim \textit{et al.}  [BESIII Collaboration],
  Phys.\ Rev.\ D \textbf{89} (2014) 052001
  [arXiv:1401.3083 [hep-ex]].
  
\bibitem{Stern:1993rg}
  J.~Stern, H.~Sazdjian and N.~H.~Fuchs,
  Phys.\ Rev.\ D \textbf{47} (1993) 3814
  [hep-ph/9301244].

\bibitem{Knecht:1995tr}
  M.~Knecht, B.~Moussallam, J.~Stern and N.~H.~Fuchs,
  Nucl.\ Phys.\ B \textbf{457} (1995) 513
  [hep-ph/9507319].

\bibitem{Ananthanarayan:2000cp}
  B.~Ananthanarayan and P.~B\"uttiker,
  Eur.\ Phys.\ J.\ C \textbf{19} (2001) 517
  [hep-ph/0012023].
  
\bibitem{Zdrahal:2008bd}
  M.~Zdr\'ahal and J.~Novotn\'y,
  Phys.\ Rev.\ D \textbf{78} (2008) 116016
  [arXiv:0806.4529 [hep-ph]].
  
\bibitem{Stoffer:2014rka}
  P.~Stoffer,
  arXiv:1412.5171 [hep-ph].
  
\bibitem{Caprini:2011ky}
  I.~Caprini, G.~Colangelo and H.~Leutwyler,
  Eur.\ Phys.\ J.\ C \textbf{72} (2012) 1860
  [arXiv:1111.7160 [hep-ph]].

\bibitem{CapriniWIP}
  I.~Caprini, G.~Colangelo and H.~Leutwyler,
  in preparation.
  
\bibitem{GarciaMartin:2011cn}
  R.~Garc\'ia-Mart\'in, R.~Kami\'nski, J.~R.~Pel\'aez, J.~Ruiz de Elvira and F.~J.~Yndur\'ain,
  Phys.\ Rev.\ D \textbf{83} (2011) 074004
  [arXiv:1102.2183 [hep-ph]].
  
\bibitem{piK}
  P.~B\"uttiker, S.~Descotes-Genon and B.~Moussallam,
  Eur.\ Phys.\ J.\ C \textbf{33} (2004) 409
  [hep-ph/0310283].

\bibitem{Pelaez:2016tgi}
  J.~R.~Pel\'aez and A.~Rodas,
  Phys.\ Rev.\ D \textbf{93} (2016) 074025
  [arXiv:1602.08404 [hep-ph]].

\bibitem{Mai:2017vot}
  M.~Mai, B.~Hu, M.~D\"oring, A.~Pilloni and A.~Szczepaniak,
  Eur.\ Phys.\ J.\ A {\bf 53} (2017) 177
  [arXiv:1706.06118 [nucl-th]].

\bibitem{Jamin:2001zq}
  M.~Jamin, J.~A.~Oller and A.~Pich,
  Nucl.\ Phys.\ B \textbf{622} (2002) 279
  [hep-ph/0110193].

\bibitem{Edera:2005dk}
  L.~Edera and M.~R.~Pennington,
  Phys.\ Lett.\ B \textbf{623} (2005) 55
  [hep-ph/0506117].

\bibitem{Moussallam:2007qc}
  B.~Moussallam,
  Eur.\ Phys.\ J.\ C \textbf{53} (2008) 401
  [arXiv:0710.0548 [hep-ph]].
 
\bibitem{Gasser:1983yg}
  J.~Gasser and H.~Leutwyler,
  Annals Phys.\  {\bf 158} (1984) 142.

\bibitem{Colangelo:2001df}
  G.~Colangelo, J.~Gasser and H.~Leutwyler,
  Nucl.\ Phys.\ B {\bf 603} (2001) 125
  [hep-ph/0103088].

\bibitem{Aston:1987ir}
  D.~Aston {\it et al.},
  Nucl.\ Phys.\ B {\bf 296} (1988) 493.
 
\bibitem{Watson:1954uc}
  K.~M.~Watson,
  Phys.\ Rev.\  {\bf 95} (1954) 228.

\bibitem{Guo:2015kla}
  P.~Guo,
  Mod.\ Phys.\ Lett.\ A {\bf 31} (2016) 1650058
  [arXiv:1506.00042 [hep-ph]].

\end{thebibliography}
\end{document}